# Photonics of fungal cell wall

**Lautaro BARÓ[a], Andrés E. DOLINKO[a,c,e], Sonia ROSENFELDT[a], Diana C. SKIGIN[b,d], Cecilia CARMARÁN[a,c,*]**

[a]Universidad de Buenos Aires, Facultad de Ciencias Exactas y Naturales, Departamento de Biodiversidad y Biología Experimental, Buenos Aires, Argentina.
[b]Universidad de Buenos Aires, Facultad de Ciencias Exactas y Naturales, Departamento de Física, Grupo de Electromagnetismo Aplicado, Buenos Aires, Argentina.
[c]CONICET – Universidad de Buenos Aires, Instituto de Micología y Botánica (IINMBIO), Buenos Aires, Argentina
[d]CONICET- Universidad de Buenos Aires, Instituto de Física (IFIBA), Buenos Aires, Argentina
[e]Universidad de Buenos Aires, Facultad de Agronomía, Departamento de Ingeniería Agrícola y Uso del Suelo, Catedra de Física, Buenos Aires, Argentina.

* Corresponding author: Current address: Facultad de Ciencias Exactas y Naturales, Depto. de Biodiversidad y Biología Experimental, Av. Intendente Güiraldes 2160 - Ciudad Universitaria, Piso 4. C1428EGA, Buenos Aires, Argentina.Tel.: (+54 11) 4787 2706.
E-mail address: carmaran@bg.fcen.uba.ar (C. Carmarán)

**Abstract**

The cell wall serves as a mechanical protection for the fungi and controls the traffic of molecules entering the cell. However, the optical role of the cell wall has not been fully investigated. In this work, we use a computational simulation to evaluate the optical response of the fungal cell wall. The main advantage of this approach is that the scattering structure can be introduced within the simulation via transmission electronic microscope (TEM) images of its cross section which in this case correspond to real fungal cell walls. The obtained results indicate that the reflectivity of the fungal cell wall depends on the incident wavelength, and then, it modulates the light that reaches the interior of the cell. This suggests that the light stimulus that reaches the inner part of the cell could depend on the structural characteristics of the cell wall. Additionally, regarding the analysis of color, our finding opens up a new and interesting approach to investigate the ecological role of UV light in relation to fungal organisms and their microhabitat.

## 1. Introduction

Survival, dispersal, growth and reproduction of fungal organisms can be strongly influenced by light (Rodriguez-Romero et al. 2010; Menezes et al. 2015; Fuller et al. 2016; among others). The response of fungi to these light stimuli may take different forms, and depends on time exposure and light characteristics such as wavelength and intensity of the incident radiation that reaches the cells (Fuller et al. 2015). Several works have assessed the effects of light on the cellular processes, the pathways involved have been explored in some species, and even light receptors and their genomic bases have been described (Fuller et al. 2015). More than one hundred fungal species have been evaluated, and different spectral responses have been reported (Marsh et al.1959; Purschwitz et

al.2006; Herrera-Estrella & Horwitz 2007). Light receptors have been described as components of the plasmatic membrane (opsins) or the cytoplasm (Rodriguez-Romero et al. 2010). Additionally, proteins localized in the nucleus (WC1 and WC2) have been related with the blue light perception (Schwerdtfeger & Linden 2001; Yang et al. 2016). However, in the investigations on the cellular events related to light in fungal organisms, few mentions (Fuller et al. 2013) have been made about the role of the first obstacle that the cell presents to light: the cell wall.

The cell wall of filamentous fungi is mainly composed of glucans (30–80%), chitin and chitosan (1–15%), mannans and/or galactomannans, and glycoproteins (Aranda-Martinez et al. 2016). This structure is essential to maintain the cell shape, but the fungal cell wall also appears associated to developmental, physiological, and biochemical processes allowing cell growth, division, and the formation of various cell types (Aranda & Martinez et al. 2016; Liesche et al. 2015). The relevance of this structure lies in the genomics information, since up to 20% of the genome can be linked to the composition and the structure of the cell wall (Ruiz Herrera 2016). Some studies about the interaction between cells and light, specially in unicelullar algae, were reported (Bhowmik & Pilon 2016; Drezek et al. 1999). However, to the best of our knowledge, the influence of the fungal cell wall has not been analysed from a photonical point of view.

In this paper we explore the influence of the fungal cell wall as a modulator of the light that reaches the inner part of the cell, by considering it as a photonic structure. We analyse the dependence of its optical response with the wavelength in order to evaluate the transmitted spectrum that reaches the interior of the cell. Taking into account the limitations involved in the evaluation of the optical response by experimental methods, in this investigation we use a photonic simulation method specially designed to compute the optical response of photonic biological structures (Dolinko 2009; Dolinko et al.2012; Dolinko & Skigin 2013). This method permits us to obtain the optical response of a complex microstructure imaged via a digitalized TEM micrography. Using this method, it was showed that the optical response of the corrugated pellicle covering certain unicellular algae called Euglenoids prevents the penetration of UV radiation by means of a photonic mechanism (Inchaussandague et al. 2017). The aim of this research is to evaluate and describe the optical response of the fungal cell wall by using analyses *in silico* on fungal cells from mycelia developed in both light and dark conditions.

**2. Materials and methods**

The analyses were carried out on native Argentinean strains of *Inonotus rickii* (BAFC cult 4302) and *Peroneutypa scoparia* (BAFC 3010). The strains were inoculated in 90 mm Petri plates (20 ml medium per plate) containing malt extract (12.7 g/l), glucose (10 g/l) and agar (20 g/l) (MEA). The inoculated plates were incubated at 24 °C for 15 days. Four plates were grown in dark, and four plates were maintained in 24hs light.

From each plate 8 samples of mycelia were taken using a punch. The samples were pre-fixed in 2.5 % glutaraldehyde in phosphate buffer (pH 7.2) for 2 hs, then post-fixedin OsO4 at 28 ºC in the same buffer for 3 hs followed by dehydration in an ethanol series and embedded in Spurr's resin. Thin sections were made on a Sorval ultramicrotome, then stained with uranyl acetate and lead citrate. The sections were observed and photographed with three different transmission electron microscopes: JEOL-JEM1200 EX II TEM at 85.0 kV, TEM PHILIPS EM 301, and TEM Zeiss

EM109T in order to minimize the variations introduced by each microscope. The samples were also observed with an Olympus BX60 M Bright field reflected light microscope.

**2.1 Photonic simulation method**

The Photonic simulation method reproduces the propagation of electromagnetic waves by an analogy with transverse mechanical waves in a planar network of particles coupled by elastic springs (Dolinko 2009). This approach permits simulating optical phenomena involving dielectric materials with arbitrary distributions of refraction index illuminated by transverse electric (TE) polarized light in spaces with translational symmetry (Dolinko et al. 2012; Dolinko & Skigin 2013). One of the key features of this method is the possibility of defining the simulation domain by means of digital images or bitmaps. Each pixel in the image represents the position of the particle in the array. In this manner, a digital image automatically defines the size of the simulation domain. An appropriate scale given in [nm/pixel] links the size of an object in the digital image, in pixels, with the actual physical size of the sample to be investigated, which in the case of photonic structures is usually measured in the scale of nanometers.

**2.2 Evaluation of the cell wall optical response**

The photonic simulation was used to obtain the optical response of sections of hypha from TEM images of the samples. To define the refraction index distribution associated with these images, the grey levels of each bitmap are assigned to a given refraction index interval of known mean value. According to the values reported in the literature, in the present work we have used a mean value of 1.5 for the cell wall and 1.34 for the internal cell medium (Woolley 1975). Taking into account that the cell wall is almost perfectly transparent, in this work we neglect absorption by the structure.

The figure 1 shows the geometric location of the cell wall in relation to the incident light beam in the simulation. It is also showed the angle $\alpha$ of incidence. Fig. 1 also shows schematically the reflected and transmitted light from the structure. The numerical analysis was carried out on the TEM photographs in which the cell wall section appears complete and without damage. A total of 40 fragments of cell wall have been evaluated with the simulation, i.e.,10 replicas of each of the four cases studied: *Inonotus rickii* and *Peroneutypa scoparia* grown in continuous darkness (24 hours), and in continuous light (24 hours). For each cell wall section, the reflected spectrum was obtained as a function of the wavelength ($\lambda$) and the observation angle ($\alpha$), which also permits to estimate the color observed at each observation direction taking into account the spectral sensitivity of the human eye (Inchaussandague et al, 2010). This permits to reproduce the iridescence (variation of the color with the observation direction) produced by the illuminated cell wall. The procedure to obtain the observed color combines the spectrum of the illuminating source, the reflectivity spectrum of the object in the visible range (380nm to 780nm) and the three standard primaries, the CIE X, Y, and Z tristimulus values of the human eye chromatic sensitivity (Lozano 1978). The spectrum of the illuminant $D(\lambda)$ was taken here as a constant equal to the unity, i.e. perfect white light.

The grey levels were used to assign the refraction index, using an adequate linear function. The value of the grey levels itself in the TEM image does not affect the results of the simulation, since we use the distribution of grey levels in the image and not their values, to associate the refraction index in each region of the image. This simulation procedure was previously validated (Dolinko, 2009) and used as a method for determining the optical response of highly complex biological photonic structures (Dolinko and Skigin, 2013) and applied for a photonics study of Myxomycetes using their TEM images (see Dolinko et al. 2012)

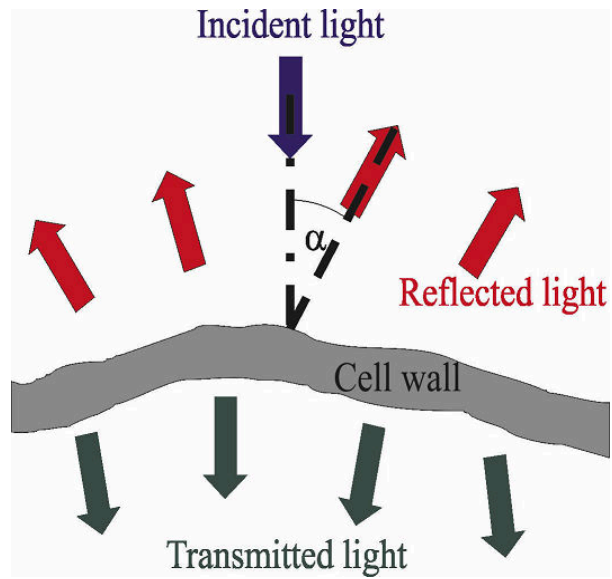

Figure 1 – Schematic model of the cell wall in relation to the incident, reflected and transmitted light. $\alpha$ is the observation angle.

To reveal the iridescence in the UV range, a similar procedure was carried out. Since this region of the spectrum cannot be perceived by the human eye, the obtained hues are computed in a false color scale. The reflectance spectrum in theinterval 250nm-390nm corresponding to the ultraviolet range was transported to the range 390nm-530nm (i.e. the violet-blue range of the visible spectrum), and then used to calculate the color. In this manner, the farther ultraviolet wavelengths are visualized in violet color and the near ultraviolet wavelengths in blue.

The total reflectance of each sample was also calculated by integration of the reflected spectra over the whole range of observation angles.

### 3. Results

Results of the analysis performed on the studied organisms are summarized in Fig. 2. Each panel of this figure shows a TEM image of the cell wall introduced in the simulation (upper part), and the reflectance as a function of the optical wavelength ($\lambda$) and the observation angle ($\alpha$) (lower part). The right color scale corresponds to the reflectance values, which range between 0 and $3\times10^{-3}$ for all the examples. We also show the computed colors for each observation angle, for the visible (right color bar) and the UV range (false color, left color bar).

The cell wall of *Inonotus rickii* presents, in most cases, three layers of different electrodensity for both conditions of growth (darkness / light), and show a total wall thickness that ranges between 60 and 150 nm. The thickness of the cell wall (darkness / light) of *Peroneutypa scoparia* ranges between 90 and 240 nm, and also three layers can be observed.

Regarding the optical response of the cell wall, all the samples studied present the typical response expected for a curved inhomogeneous thin film, independently of the species and growth conditions. It can be observed that the reflected light is more widely scattered from the samples of larger curvature. The light is mainly reflected along the normal direction (α=0, see Fig. 1), although a fraction, which depends on the sample, is scattered in different directions depending on the wavelength. It can be observed that, as expected, the spreading of the reflected spectrum increases with the wavelength.

Figures 2(a) and 2(d) show the results for the strains of *Inonotus rickii*: panels (a) and (b) correspond to specimens grown under light and (c) and (d) to those grown under darkness. It can be observed that in all these cases, the obtained reflected color is mainly red. There are not substantial differences between the samples grown under different conditions of illumination. In contrast, for the strains of *Peroneutypa scoparia*, (Figs. 2(e)-2(h)), the blues prevail in the samples developed in light (2(e)-2(f)) and the greens in those developed in darkness (2(g)-2(h)). It is also observed that in these species the obtained color is asymmetric in relation to the observation angle. According to the computed false color in the UV range, a certain degree of iridescence is also noticed in this wavelengths range.

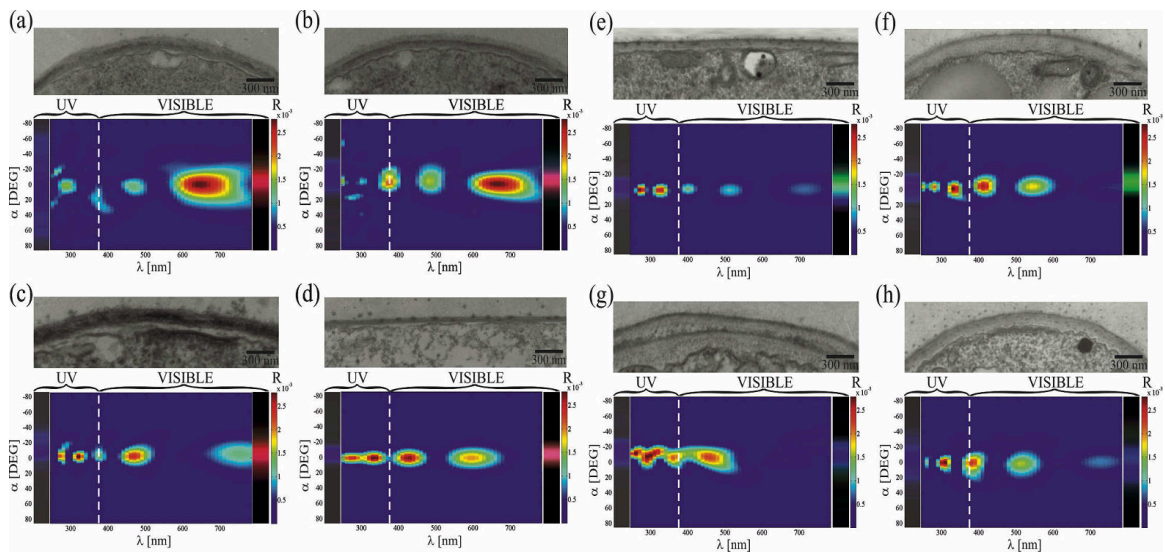

Figure 2 – Simulated spectrum as a function of the optical wavelength (λ) and observation angle (α). (a,b) *Inonotus rickii* grown under darkness; and (c,d) *Inonotus rickii* grown under light; (e,f) *Peroneutypa scoparia* grown under darkness; and (g,h) *Peroneutypa scoparia* grown under light. The color bars at each side of the spectrum show the resulting color as a function of the observation angle for UV (in false color, left) and for the visible range (right). The right color bar corresponds to the reflectance magnitude (R) plotted in the showed spectra.

Figure 3 (a-h) shows the reflectance integrated over the observation angles range (R), associated with each of the analysed TEM images. These curves show that the analysed sections of cell wall exhibit a set of maxima, whose spectral distance increases with the wavelength, as expected for a homogeneous thin film (Kinoshita 2008). In the case of *Inonotus,* the distance between peaks seems to be smaller for those samples grown under constant light (2c, 2d) than those grown under darkness (2a, 2b). However, no evident dependency of the reflectance spectra of *P. scoparia* with the growth conditions is observed.

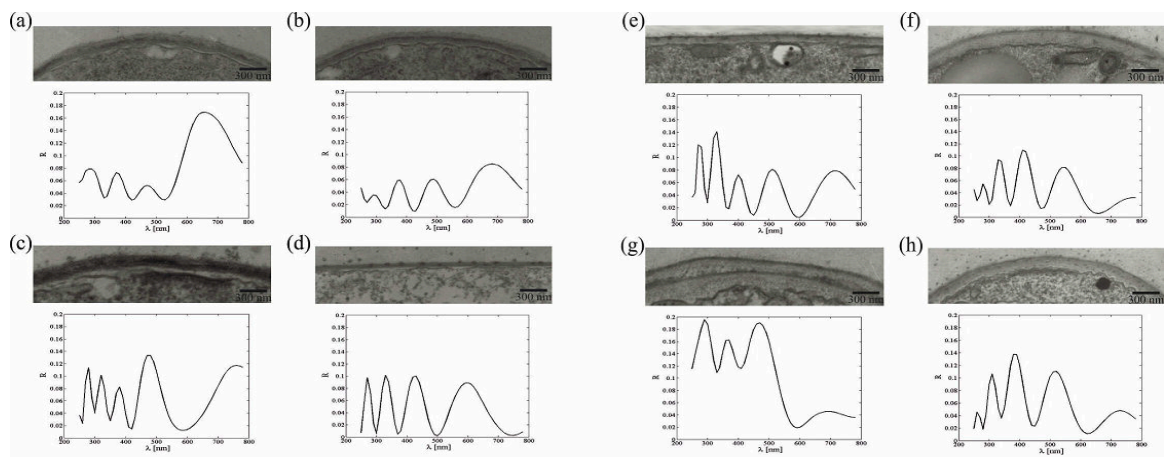

Figure 3 – Simulated spectrum as a function of the optical wavelength and integrated over the observation angle (a,b) *Inonotus rickii* grown under darkness; and (c,d) *Inonotus rickii* grown under light; (e,f) *Peroneutypa scoparia* grown under darkness; and (g,h) *Peroneutypa scoparia* grown under light.

Additional computational analyses on TEM images of different species of fungi previously published by other authors were also carried out. The results obtained are showed in Figure 4. In the particular case of TEM photographs of *Botrytis cinerea,* the analysis was performed on images of the strain Bc05.10 and of the altered cell wall of the DBcphy3-P21 mutant (Hu et al. 2014). Our results show that in all the cases studied, *Candidaalbicans*, *Paracoccidioides brasiliensis* and *Botrytis cinerea*, the light spectrum that reaches the interior of the cell is modullated by the fungal cell walls by means of a photonic mechanism.

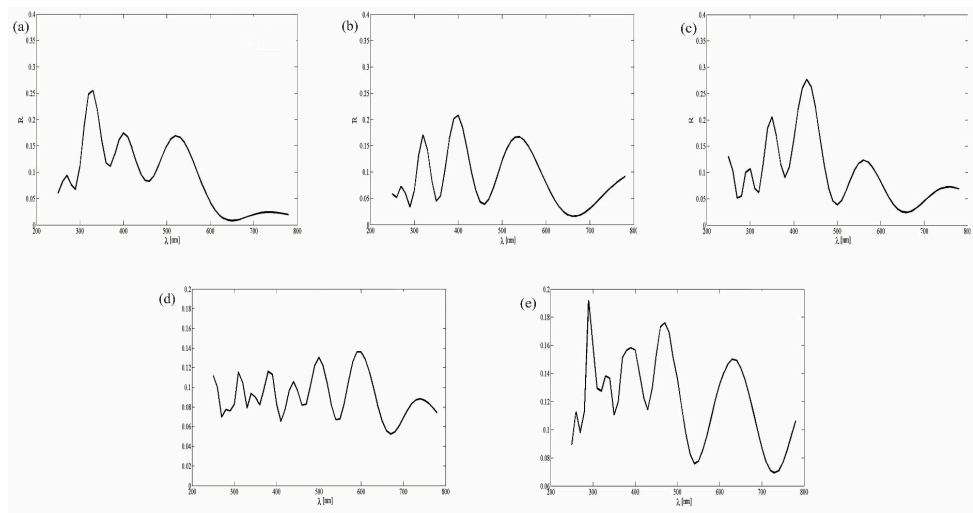

Figure 4 – Simulated spectrum as a function of the optical wavelength and integrated over the observation angle (a) *Candida albicans* (Alvarez et al. 2008 Fig. 6A); (b) *Paracoccidioides brasiliensis* (Lopera et al. 2008 Fig. 3B); (c) *Paracoccidioides brasiliensis* (Lopera et al. 2008 Fig. 3D); (d) *Botrytis cinerea* -Bc05.10 strain- (Hu et al. 2014 Fig. 6C); and (e) *Botrytis cinerea* -DBcphy3 mutant- (Hu et al. 2014 Fig. 6C).

## 4. Discussion

The photonic simulation used here is suitable for dealing with dielectric objects of translational symmetry with arbitrary spatial refraction index distributions and shapes (Dolinko 2009; Dolinko & Skigin 2013). In particular, it has been very useful for investigating the electromagnetic response of biological photonic structures that usually present a high degree of complexity in their geometry as well as in the materials involved (Dolinko et al. 2012). Based on this approach, our findings indicate that the fungal cell wall behaves as a photonic structure that presents an optical response similar to that of an inhomogeneous thin film. The inhomogeneities of this film are manifested in the obtained spectra as slight deviations from the response of an homogeneous planar slab of similar width and average refraction index.

The ultrastructure observed in the TEM images corresponds to the general description of several species (Ruiz Herrera 2016), which shows several layers with different electrodensity.

Among the analyses performed *in silico,* stronger reflectance in *I. rickii* specimens was observed in the red region of the spectrum for the tests developed in darkness, while the reflectance is slightly increased in the UV region for the specimens grown in light. This differentiated behaviour are associated to the different electrodensity observed in the layers of the cell wall for the specimens grown in different light conditions. These responses could be species-specific, taking into account that *Peroneutypa scoparia* does not exhibit any dependency of the electromagnetic response with the light growth conditions.

Similar results were obtained for previously published photographs. The findings presented here indicate that the spectral composition of the light that reaches the inner part of the cell is modified by the cell wall by means of a photonic mechanism. The reflectance observed reaches a maximun of 30 % in some cases. Hu et al. (2014) suggest that in *Botrytis cinerea* some alterations of processes mediated by light, like those mediated by phytochrome like HK gene Bcphy3, may be related to the modification of the cell wall structure, giving support to the hypothesis of an indirect effect of Bcphy3 on cell wall integrity. Taking into account the vegetative development of fungal mycelia in the substrate, where they grow in several layers, the cumulative effect of reflected energy could drive to the decrease of the UV radiation reaching the interior of the cells. Important implications of our study derive from the consequences of the mycelia structure, which developes in several layers when it grows superficially. Our findings indicate that a total absorbance of at least a narrow spectral bandwidth is to expect a few microns below the surface. This phenomenon could represent a protection strategy in the case of fungi without additional protecting agents like melanin (Ruiz Herrera 2012). More studies (currently in progress) are necessary to investigate this hypothesis.

Additionaly, the analysis of color opens up a new and interesting approach to investigate the ecological role of UV light in relation to fungal organisms and their microhabitat. It is true that not always reflection of UV light necessarily implies a communicative role and an adaptive feature. However, taking into account that many animal species are able to perceive ultraviolet wavelengths (Cronin & Bok 2016), the obtained results indicate that some groups of arthropods could have some level of perception of mycelia. This ability could be increased by the selective reflection of light of the cell wall demonstrated in this work. In this context, it might be interesting to explore the role of the UV reflectance in fungal organisms.

## 4. Conclusions

This work explores for the first time the photonic properties of the fungi cell wall suggesting that the role of the cell wall as a light modulator should be included as an additional factor in studies relative to the light and its effects on cellular processes. Additionally, it opens up new perspectives on ecological aspects of the microcosmos.

**Authors' contributions**
Setting up the computational model: A D. Lab work and acquisition of data:L.B, A. D.Data analysis: all. Conception and design of the investigation and work: A.D., C.C. Drafting the manuscript: all. Revising it critically for important intellectual content and final approval of the version to be published: all. Agreement to be accountable for all aspects of the work in ensuring that questions related to the accuracy or integrity of any part of the work are appropriately investigated and resolved: all.

**Competing interests**
The authors declare that the research was conducted in the absence of any commercial or financial relationships that could be construed as a potential conflict of interest.

**Funding.**
This work was funded by Universidad de Buenos Aires (UBACyT 20020150100028BA, UBACyT) and Consejo Nacional de Investigaciones Científicas y Técnicas (CONICET) (PIP 112-201101-00451, PIP 11220150100956).

**Data Availability**

The computational simulation used in this article have been previously published, the corresponding references are indicated in Materials and Methods.

**References**

Alvarez FJ, Douglas LM, Rosebrock A, Konopka JB. 2008 The Sur7 protein regulates plasma membrane organization and prevents intracellular cell wall growth in *Candida albicans*. *Molecular Biology of the Cell* **19,** 5214–5225.

Aranda-Martinez, A, Lopez-Moya, F, & Lopez-Llorca, LV. 2016 Cell wall composition plays a key role on sensitivity of filamentous fungi to chitosan. *Journal of basic microbiology* **56,** 1059-1070.

Berthier S, 2007 *Iridescences, the physical colours of insects*. Springer, New York, USA.

Bhowmik A, Pilon L, 2016 Can spherical eukaryotic microalgae cells be treated as optically homogeneous? *Journal of the Optical Sociaty of America A* **33**, 1495–1503.

Cronin TW, Bok MJ. 2016 Photoreception and vision in the ultraviolet. *Journal of Experimental Biology* **219***,* 2790–2801.

Dolinko AE. 2009 From Newton's second law to Huygens's principle: Visualizing waves in a large array of masses joined by springs. *European Journal of Physics* **30**, 1217–1228. Doi: 10.1088/0143-0807/30/6/002.

Dolinko A, Skigin D, Inchaussandague M, Carmaran C. 2012 Photonic simulation method applied to the study of structural color in Myxomycetes. *Optics Express* **20**, 15139–15148.

Dolinko AE, Skigin DC. 2013 Enhanced method for determining the optical response of highly complex photonic structures of biological origin. *Journal of the Optical Society of America A* **30**, 1746–1759. doi.org/10.1364/JOSAA.30.001746

Drezek R, Dunn A, Richards-Kortum R. 1999 Light scattering from cells: finite-difference time-domain simulations and goniometric measurements. *Applied Optics* **38**, 3651–3661.

Fuller KK, Ringelberg CS, Loros JJ, Dunlap JC. 2013 The fungal pathogen Aspergillus fumigatus regulates growth, metabolism, and stress resistance in response to light. *MBio* **4**, e00142–13.

Fuller KK, Loros JJ, Dunlap JC. 2015 Fungal photobiology: visible light as a signal for stress, space and time. *Current genetics* **61**, 275–288.

Fuller KK, Dunlap JC, Loros JJ. 2016 Fungal light sensing at the bench and beyond. *Advances in Genetics* **96**, 1–51.

Herrera-Estrella A & Horwitz BA. 2007 Looking through the eyes of fungi: molecular genetics of photoreception. *Molecular microbiology* **64**, 5–15.

Hu Y, He J, Wang Y, Zhu P, Zhang C, Lu R, Xu L. 2014 Disruption of a phytochrome-like histidine kinase gene by homologous recombination leads to a significant reduction in vegetative growth, sclerotia production, and the pathogenicity of *Botrytis cinerea. Physiological and Molecular Plant Pathology* **85**, 25–33.

Inchaussandague M, Skigin D, Carmaran C, Rosenfeldt S. 2010 Structural color in Myxomycetes. *Optics Express* **18**, 16055–16063.

Inchaussandague M, Skigin D, Dolinko A, 2017. Optical function of the finite-thickness corrugated pellicle of Euglenoids. *Applied Optics* **56**, 5112–5120.

Kinoshita S. 2008 Structural colors in the realm of nature, (World Scientific Publishing Co., Singapore).

Liesche J, Marek M, Günther-Pomorski T. 2015 Cell wall staining with Trypan blue enables quantitative analysis of morphological changes in yeast cells. *Frontiers in Microbiology* **6**, 107.

Lopera D, Aristizabal BH, Restrepo A, Cano LE, González A. 2008 Lysozyme plays a dual role against the dimorphic fungus *Paracoccidioides brasiliensis*. *Revista do Instituto de Medicina Tropical de São Paulo.* **50**, 169–175.

Lozano R, 1978. El color y su medición, (Americalee Ed., Argentina).

Marsh PB, Taylor EE, Bassler LM. 1959 Guide to the literature on certain effects of light on fungi. Plant disease reporter; suppl. 261.

Menezes HD, MassolaJr NS, Flint SD, Silva Jr GJ, Bachmann L, Rangel DE, Braga GU. 2015 Growth under visible light increases Conidia and Mucilage production and tolerance to UV-B radiation in the plant pathogenic fungus *Colletotrichum acutatum*. *Photochemistry and Photobiology* **91**, 397–402.

Rodriguez-Romero J, Hedtke M, Kastner C, Müller S, Fischer R. 2010 Fungi, hidden in soil or up in the air: light makes a difference. *Annual Review of Microbiology* **64**, 585–610.

Ruiz-Herrera J. 2016 Fungal Cell Wall: Structure, Synthesis, and Assembly. CRC Press; 2 edition.

Purschwitz J, Müller S, Kastner C, Fischer R. 2006 Seeing the rainbow: light sensing in fungi. *Current opinion in microbiology* **9**, 566–571.

Schwerdtfeger C, Linden H. 2001 Blue light adaptation and desensitization of light signal transduction in Neurosporacrassa. *Molecular microbiology* **39**, 1080–1087.

Woolley JT, 1975. Refractive index of soybean leaf cell walls. Plant Physiology, **55**, 172-174.

Yang T, Guo M, Yang H, Guo S, Dong C. 2016 The blue-light receptor CmWC-1 mediates fruit body development and secondary metabolism in *Cordyceps militaris*. *Applied microbiology and biotechnology* **100**, 743–755.